\documentclass[sigconf, nonacm]{acmart} 
\AtBeginDocument{%
  }

\copyrightyear{2026}
\acmYear{2026}
\setcopyright{cc}
\setcctype{by}
\acmConference[C\&C '26]{Creativity and Cognition}{July 13--16, 2026}{London, United Kingdom}
\acmBooktitle{Creativity and Cognition (C\&C '26), July 13--16, 2026, London, United Kingdom}
\acmDOI{10.1145/3803784.3807565}
\acmISBN{979-8-4007-2583-8/2026/07}

\usepackage{listings}

\begin{document}

\title{AttentionBender: Manipulating Cross-Attention in Video Diffusion Transformers as a Creative Probe}


\author{Adam Cole}
\email{a.cole@arts.ac.uk}
\orcid{0000-0001-9715-314X}
\affiliation{%
  \institution{University of the Arts London}
  \city{London}
  \country{UK}
}

\author{Mick Grierson}
\email{m.grierson@arts.ac.uk}
\orcid{0000-0002-6981-5414}
\affiliation{%
  \institution{University of the Arts London}
  \city{London}
  \country{UK}
}



\begin{abstract}
We present \textit{AttentionBender}, a tool that manipulates cross-attention in Video Diffusion Transformers to help artists probe the internal mechanics of black-box video generation. While generative outputs are increasingly realistic, prompt-only control limits artists’ ability to build intuition for the model’s material process or to work beyond its default tendencies. Using an autobiographical research-through-design approach, we built on Network Bending to design \textit{AttentionBender}, which applies 2D transforms (rotation, scaling, translation, etc.) to cross-attention maps to modulate generation. We assess \textit{AttentionBender} by visualizing 4,500+ video generations across prompts, operations, and layer targets. Our results suggest that cross-attention is highly entangled: targeted manipulations often resist clean, localized control, producing distributed distortions and glitch aesthetics over linear edits. \textit{AttentionBender} contributes a tool that  functions both as an Explainable AI style probe of transformer attention mechanisms, and as a creative technique for producing novel aesthetics beyond the model’s learned representational space.
\end{abstract}



\begin{teaserfigure}
  \includegraphics[width=\linewidth]{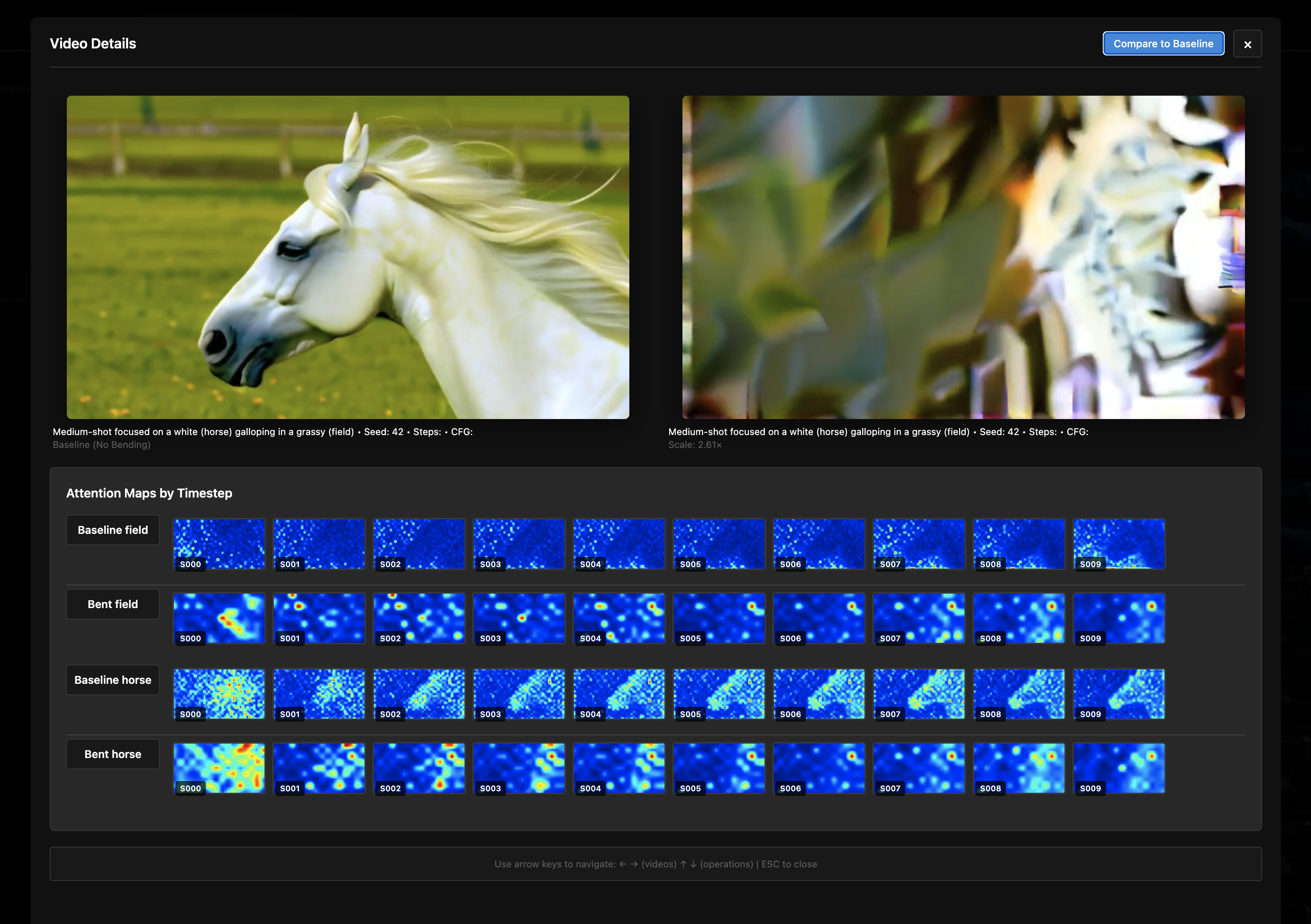}
  \caption{\textbf{Probing the internal material of generative video with \textit{AttentionBender}}. The system enables artists to intervene directly in the cross-attention mechanism of a Video Diffusion Transformer. Here, a standard generation (top-left) is compared against a "bent" variation (top-right) where the attention map for the subject has been spatially scaled. The bottom timeline visualizes the internal attention dynamics step-by-step, exposing how the mathematical intervention propagates into the final aesthetic result.}
  \label{fig:video_details}
\end{teaserfigure}

\received{05 February 2026}

\maketitle
\section{Introduction}

As AI tools continue to proliferate, a striking question remains: do artists understand the inherent properties of their material, or are they working at the mercy of opaque black box operations? Generative models are increasingly understood within creative practice as a medium defined by specific constraints, affordances, and aesthetic possibilities. In this view, the model is not only a production tool, but a material system that rewards sustained engagement and the gradual formation of tacit craft knowledge \cite{hemmentExperientialAIArts2024, broadUsingGenerativeAI2024}. Recent HCI work on creative communities using generative models also shows that artists actively develop informal mental models, folk theories, and workflows to make sense of how these systems behave, especially when they are difficult to predict or control \cite{shelbyGenerativeAICreative2024}.

However, many contemporary generative AI tools suffer from an interpretability paradox \cite{burrellHowMachineThinks2016, schneiderExplainableGenerativeAI2024}. As outputs become more realistic and better aligned with textual prompts, these systems can appear increasingly legible at the surface level. Yet the dominant prompt-based web interfaces still reveal little about the internal generative process, which remains invisible and remote \cite{zamfirescu-pereiraWhyJohnnyCant2023, torricelliRoleInterfaceDesign2024}. This interaction model limits the depth of intuition artists can cultivate for the material process itself, and it tends to confine exploration to whatever the interface makes available and convenient \cite{torricelliRoleInterfaceDesign2024}. The growing literature on prompt engineering reflects this dynamic: prompting becomes a learned creative skill, but it remains an indirect way of working, where localized intent and mechanism-level understanding are difficult to achieve through language alone \cite{liuDesignGuidelinesPrompt2022, oppenlaenderPromptingAIArt2025}.

In experimental film and media arts, there is a long tradition of artists and engineers manipulating, distorting, and otherwise interfering with technical apparatuses in order to gain a more intimate knowledge of their tools \cite{youngbloodExpandedCinema1970, ghazalaCircuitBendingBuildYour2005, collinsHandmadeElectronicMusic2006, menkmanGlitchMomentum2011}. This orientation also has a clear methodological lineage in HCI and critical computing. Critical technical practice frames technical work as a mode of inquiry, where intervening in a system can surface its assumptions and structure \cite{agreCriticalTechnicalPractice1997}. Critical making similarly treats hands-on construction and intervention as a way of producing knowledge through practice, reflection, and articulation \cite{rattoCriticalMakingConceptual2011}. In creative machine learning, this tradition becomes explicit through Network Bending, which inserts deterministic transformations into trained generative networks at inference time to reveal the network's internal mechanics and open new expressive possibilities \cite{broadNetworkBendingExpressive2021}.

This material, interventionist approach intersects with a growing discourse around Explainable AI for the Arts (XAIxArts) \cite{bryan-kinnsExplainableAIArts2023, bryan-kinnsXAIxArtsManifestoExplainable2025}, which argues that explainability should be evaluated by how it supports artist agency, reflection, and creative autonomy, rather than only by fidelity to a single “ground truth” explanation . Within this framing, explainability is not only a descriptive report about a model, but something that can be enacted through tools that allow artists to probe, test, and reconfigure internal processes \cite{bryan-kinnsExploringXAIArts2023}. Recent work on “explainability-in-action” pushes in this direction by treating interactive intervention as a meaningful route to interpretability in generative systems \cite{abuzuraiqExplainabilityinActionEnablingExpressive2025}.

Until now, creative network bending interventions have been developed primarily for GAN-based pipelines \cite{broadNetworkBendingExpressive2021, aldegheriHackingGenerativeModels2023}, such as StyleGAN \cite{karrasStyleBasedGeneratorArchitecture2019a}, UNet-based image diffusion models \cite{dzwonczykNetworkBendingDiffusion2024, abuzuraiqExplainabilityinActionEnablingExpressive2025}, such as Stable Diffusion \cite{rombachHighResolutionImageSynthesis2022}, and neural audio models \cite{mccallumNetworkBendingNeural2020, manzBraveDesigningEmbedded2025}, such as RAVE \cite{caillonRAVEVariationalAutoencoder2022}. However, the architectures that power state-of-the-art generative systems are shifting. Diffusion transformer (DiT) \cite{peeblesScalableDiffusionModels2023b} backbones have recently emerged as a scalable alternative to UNet-based diffusion \cite{esserScalingRectifiedFlow2024a, openaiSoraBlogpost2024, wanteamWanOpenAdvanced2025}, shifting research and production systems toward transformer-centric architectures built around self-attention and cross-attention mechanisms \cite{vaswaniAttentionAllYou2017, dosovitskiyImageWorth16x162020a}. Meanwhile, video generation is evolving quickly in parallel, and diffusion-based video generation has rapidly expanded in research and practice \cite{hoVideoDiffusionModels2022, singerMakeAVideoTexttoVideoGeneration2022, luVDTGeneralpurposeVideo2023, openaiSoraBlogpost2024, kongHunyuanVideoSystematicFramework2025, wanteamWanOpenAdvanced2025}. As these contemporary approaches to video generation proliferate, it becomes essential that artists gain hands-on knowledge of their internal mechanisms. Likewise, it is valuable to explore new mechanisms to intervene in these models' internal workings which open up novel creative possibilities.

Our work addresses both of these needs, by developing an artist-facing creative XAI approach that targets video diffusion transformers, specifically by using a network bending approach that treats cross-attention as a direct site of intervention during inference. Our approach, titled \textit{AttentionBender}, focuses on cross-attention because it is one of the more directly interpretable junctions where the text prompt conditions the visual latent \cite{hertzPrompttoPromptImageEditing2022a, cheferAttendandExciteAttentionBasedSemantic2023, tangWhatDAAMInterpreting2022a}, making it a practical place to intervene and trace downstream effects. \textit{AttentionBender} modulates the generative process by intercepting the cross-attention computation during inference and applying spatial manipulations, such as rotation, scaling, and translation, to the resulting cross-attention maps. These interventions can be targeted across token selections, timestep ranges, and layer ranges. Rather than promising surgical control, \textit{AttentionBender} is designed to support a mode of inquiry: by bending a central internal mechanism and reading its effects across systematic variation, we can build a more grounded intuition for how the model behaves as an artistic material.

The project follows an autobiographical research-through-design approach, using our creative-technical practice as the site of inquiry and tool development \cite{zimmermanResearchDesignMethod2007, neustaedterAutobiographicalDesignHCI2012}. We evaluate \textit{AttentionBender} through a systematic sweep of possible variations, visualizing over 4,500 video outputs across prompts, seeds, operations, and intervention targets. We interpret these results through comparative viewing in a custom interface designed for high-level filtering and drill-down inspection. This workflow enables pattern-level comparison across families of outputs to support intuition-building, while also opening creative possibilities beyond prompt-only workflows.

This paper makes three contributions:
\begin{enumerate}
    \item \textbf{AttentionBender:} an inference-time network bending pipeline for video diffusion transformers that enables spatial transforms of cross-attention maps with the ability to target specific prompt tokens, diffusion timesteps, and network layers.
    \item \textbf{Probing workflow and comparative visualization system:} a systematic sweep protocol and grid-based interface that connects micro-level inspection with high-level navigation, supporting both intuition-building and deliberate creative departures from the model’s typical tendencies.
    \item \textbf{Pattern-level observations from a systematic sweep:} a taxonomy of model behaviors derived from the sweep, identifying recurring tensions such as entanglement and resilience, alongside examples showing how these interventions can open new creative possibilities.
\end{enumerate}

\section{Related Works}

\subsection{Artist Practice with Generative Video}
\label{sec:artist-practice-video}

Generative video is rapidly transitioning from experimental novelty to a central instrument within communities of practice ranging from high-budget film production to independent creators \cite{zhangGenerativeAIFilm2025a, finneyAIScreenSector2025a, lyuPreliminaryExplorationYouTubers2024}. However, the current landscape of creative tools is characterized by a tension between fidelity and agency \cite{casacubertaDisembodiedCreativityGenerative2025}. The dominant commercial platforms, such as SORA \cite{openaiSora2Here2025a}, Google Veo \cite{humeMeetFlowAIpowered2025}, and RunwayML \cite{runwayRunwayResearchIntroducing2025}, market themselves on their realistic performance and world building capabilities \cite{openaiVideoGenerationModels}. Yet, their interface design relies almost exclusively on prompt-based interfaces, encouraging a "black-box" mental model where the generative process remains remote and invisible \cite{burrellHowMachineThinks2016, tredinnickBlackboxCreativityGenerative2023}. As noted in recent HCI research on creative-AI workflows, this lack of transparency limits the material understanding artists can develop, confining their agency to inputs and outputs while obscuring the generative mechanism itself \cite{rajcicDiffractiveAnalysisPromptBased2024a, torricelliRoleInterfaceDesign2024}.

While open-source ecosystems, such as ComfyUI \cite{comfy-orgComfyUIMostPowerful2023}, and models, such as Wan \cite{wanteamWanOpenAdvanced2025}, offer an alternative, they present distinct challenges. Node-based interfaces successfully visualize the data flow through a pipeline, and provide greater control mechanisms like Low Rank Adapter (LoRA) networks \cite{huLoRALowRankAdaptation2021a}, ControlNet \cite{zhangAddingConditionalControl2023}, and IPAdapters \cite{yeIPAdapterTextCompatible2023}. Despite making the pipeline visible, the core generative act (the iterative denoising process) remains hidden behind a "Sampler" node, often represented only as a progress bar. Furthermore, with standard open-source video models ranging from 1 billion to over 14 billion parameters \cite{kongHunyuanVideoSystematicFramework2025, wanteamWanOpenAdvanced2025, pengOpenSora20Training2025, yangCogVideoXTextVideoDiffusion2024}, the sheer complexity of the architecture resists meaningful inspection \cite{burrellHowMachineThinks2016, zimmermannScaleAloneDoes2023}. \textit{AttentionBender} intervenes in this gap, attempting to establish an intimacy with the internal process that prompt-based and node-based interfaces currently obscure.

\begin{figure}[h]
  \centering
  \includegraphics[width=\linewidth]{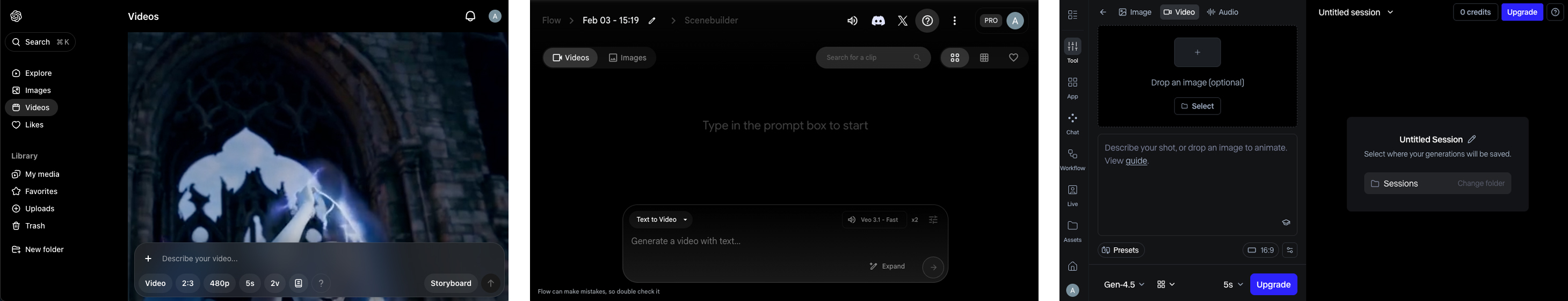}
  \caption{AI video generation user interface for three major commercial platforms as of February 1, 2025: OpenAI Sora, Google VEO Flow, and RunwayML (from left to right,). All three tools are dominated by prompt-box UIs with the generative process being totally obscured.} 
  \label{fig:full_grid}
\end{figure}

\subsection{Diffusion Transformer Architecture}

\subsubsection{Diffusion Models}
Across commercial and open-source ecosystems, diffusion models have become the dominant architecture for generative synthesis \cite{yazdaniGenerativeAIDepth2025}, outperforming Generative Adversarial Networks (GANs) in stability, fidelity, and diversity \cite{dhariwalDiffusionModelsBeat2021, yangDiffusionModelsComprehensive2024}. The core mechanism of these models is a probabilistic denoising process consisting of a forward and reverse pass \cite{hoDenoisingDiffusionProbabilistic2020}. In the forward pass, Gaussian noise is incrementally added to the data over a fixed schedule of timesteps ($T$). In the reverse pass, a neural network is trained to predict and remove this noise, iteratively reconstructing the original data distribution from pure randomness. Video diffusion extends this paradigm into the temporal domain, treating the input as a 3D volume of frames ($F$) alongside height and width, rather than a static 2D image \cite{hoVideoDiffusionModels2022}. Contemporary systems typically employ a \textit{Latent Diffusion} pipeline \cite{rombachHighResolutionImageSynthesis2022}. Rather than operating in high-dimensional pixel space, these models utilize a Variational Autoencoder (VAE) \cite{kingmaAutoEncodingVariationalBayes2013} to compress input into a lower-dimensional latent representation, $z$, where the diffusion process occurs. 

\subsubsection{Shift to Diffusion Transformers}
While early image \cite{rombachHighResolutionImageSynthesis2022} and video models \cite{hoVideoDiffusionModels2022} utilized a U-Net architecture \cite{ronnebergerUNetConvolutionalNetworks2015}, state-of-the-art systems have shifted toward the Diffusion Transformer (DiT) backbone \cite{peeblesScalableDiffusionModels2023b}. By moving beyond the spatial inductive bias of U-Nets, DiTs leverage the scalability of the transformer architecture \cite{vaswaniAttentionAllYou2017}, a design first introduced for natural language processing and subsequently brought to computer vision through Vision Transformers (ViT) \cite{dosovitskiyImageWorth16x162020}.

\subsection{The Attention Mechanism}
The fundamental operation of the transformer is the attention mechanism, which allows the model to weigh the relevance of different elements in a sequence \textit{regardless of their distance} \cite{vaswaniAttentionAllYou2017, dosovitskiyImageWorth16x162020}. This process is formalized as a retrieval operation involving three learned vectors: a \textbf{Query} ($Q$) representing the token seeking information, a \textbf{Key} ($K$) indexing the available context, and a \textbf{Value} ($V$) containing the actual feature content. These are aggregated via a scaled dot-product:
\begin{equation}
\label{eq:attention}
\text{Attention}(Q, K, V) = \text{softmax}\left(\frac{QK^T}{\sqrt{d_k}}\right)V
\end{equation}
where $\sqrt{d_k}$ is a scaling factor based on the key dimension. The output of the softmax operation forms the \textit{attention map}, a probability matrix indicating how strongly each element of the query attends to the key.

\begin{figure}[h]
  \centering
  \includegraphics[width=\linewidth]{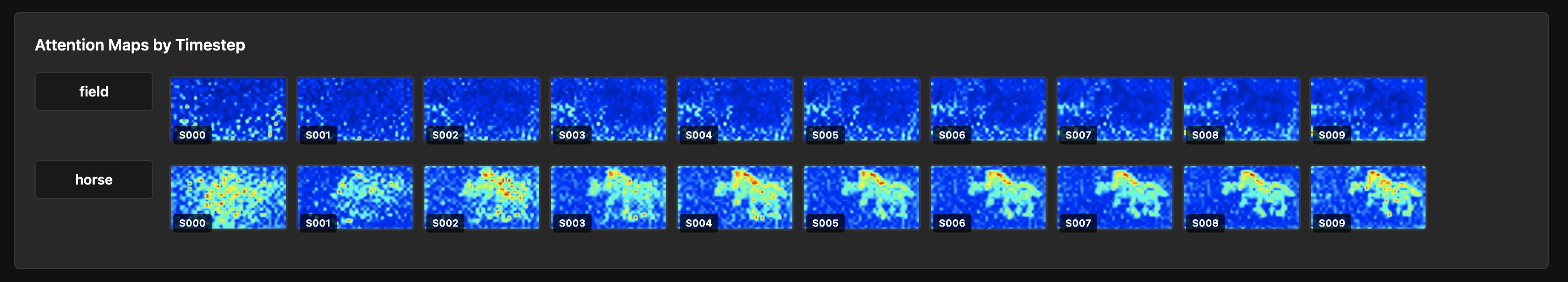}
  \caption{\textbf{Attention maps} visualized for the token "horse" and "field" for a video generated with the prompt: "white \textit{horse} galloping in a grassy \textit{field}. The brighter sections highlight areas with higher attention values. For the horse token, attention is higher where the \textit{horse} will be in the final composition, while the \textit{field} token's attention maps highlight the background. This figure shows the development of this cross-attention focus over 10 diffusion steps.} 
  \label{fig:full_grid}
\end{figure}
  
In text-to-video DiTs, two forms of attention are critical \cite{peeblesScalableDiffusionModels2023}. \textit{Self-Attention} maintains spatiotemporal consistency by relating visual patches to one another \cite{hoVideoDiffusionModels2022, arnabViViTVideoVision2021}; in this case, the queries, keys, and values are all derived from the visual latents. In contrast, \textit{Cross-Attention} serves as the primary interface for conditioning. Here, the visual latents act as queries ($Q$), while the keys ($K$) and values ($V$) are projected from prompt embeddings generated by a pre-trained language model \cite{radfordLearningTransferableVisual2021, raffelExploringLimitsTransfer2023}. This makes cross-attention the specific junction where semantic intent is mapped onto visual structure.

We identify the open-source  \textbf{Wan Video} model \cite{wanteamWanOpenAdvanced2025} as the optimal site for our intervention. Wan's 1.3B parameter model employs a pure DiT architecture composed of N transformer blocks, each containing self-attention, cross-attention, and MLP layers \cite{wanteamWanOpenAdvanced2025}. Its clean, modular structure makes it an ideal candidate for probing how text conditioning influences generation.

\subsection{Explainable AI and Attention Interpretability}

Our work is situated within the field of Explainable AI (XAI) \cite{arrietaExplainableArtificialIntelligence2019}, a diverse domain seeking to make opaque algorithmic decisions intelligible \cite{mershaExplainableArtificialIntelligence2024}. Within computer vision, significant focus has been placed on the interpretability of attention \cite{cheferTransformerInterpretabilityAttention2021}. For example, DAAM visualizes the cross-attention mechanism via attention maps \cite{tangWhatDAAMInterpreting2022a}; Attend-and-Excite amplifies cross-attention scores to prevent subject neglect \cite{cheferAttendandExciteAttentionBasedSemantic2023}; and Prompt-to-Prompt swaps cross-attention maps during generation to edit composition  \cite{hertzPrompttoPromptImageEditing2022a}.

While these works provide significant insight, they are primarily oriented toward a technical audience \cite{bryan-kinnsExplainableAIArts2023, hemmentExperientialAIArts2024}. In contrast, the emerging discourse of \textit{Explainable AI for the Arts} (XAIxArts) \cite{bryan-kinnsExplainableAIArts2023}, argues that for creative practitioners, explainability should be evaluated by how it supports agency, reflection "beyond technocentric discourses" \cite{bryan-kinnsXAIxArtsManifestoExplainable2025}. \textit{AttentionBender} addresses the legibility gap identified in \ref{sec:artist-practice-video} by situating itself within this XAIxArts framework. Moving beyond diagnostic technical metrics, we design a tool that facilitates material exploration, allowing artists to develop an embodied instinct for the generative mechanism itself \cite{casacubertaDisembodiedCreativityGenerative2025}.

\subsection{Network Bending}

The most direct methodological precedent for our system is \textit{Network Bending} \cite{broadNetworkBendingExpressive2021}. In this framework, Broad et al. demonstrated that inserting deterministic filters (scale, rotate, translate) into the layers of a trained StyleGAN2 \cite{karrasAnalyzingImprovingImage2020} model could reveal its semantic hierarchy and produce novel aesthetic artifacts. However, while Broad's original framework manipulated spatial feature maps (activations), our approach targets the \textbf{relational attention maps} of the transformer architecture.

This interventionist approach has a strong lineage in experimental media arts, echoing the material filmic interventions of Stan Brakhage \cite{brakhageMothlight1963}, the hardware hacking of Nam June Paik \cite{paikMagnetTV1965}, and the circuit bending of Reed Ghazala \cite{ghazalaCircuitBendingBuildYour2005}. Network bending has since been applied to U-Net based image diffusion \cite{dzwonczykGeneratingMusicReactive2025, abuzuraiqExplainabilityinActionEnablingExpressive2025} and neural audio generation \cite{mccallumNetworkBendingNeural2020, manzBraveDesigningEmbedded2025}. However, there is limited prior work applying these interventionist principles to the diffusion transformer architecture or to generative video. \textit{AttentionBender} addresses this gap, extending the logic of network bending into the spatiotemporal domain of the DiT to specifically probe the cross-attention mechanism.

\subsection{Research Through Design}
\label{sec:research-through-design}
This work is grounded in Research through Design (RtD) \cite{zimmermanResearchDesignMethod2007}, a methodological orientation in HCI and creative computing that treats the act of making as a primary mode of inquiry. Rather than separating design from analysis, RtD understands artifacts as sites where theoretical questions are posed, tested, and refined through use. In this view, knowledge emerges not from detached evaluation alone, but from sustained engagement with designed interventions situated within real creative practice.

Prior work has emphasized that RtD is particularly well suited to domains where the phenomena under study are experiential, tacit, or difficult to formalize through metrics alone \cite{zimmermanResearchDesignMethod2007}. Generative AI systems for artistic practice fall squarely within this category: their behavior unfolds across many iterations, their aesthetic control is complex, and meaningful understanding develops through exposure.

Within RtD, autobiographical design offers a complementary approach in which researchers draw on their own long-term, expert use of a system as a legitimate source of insight \cite{neustaedterAutobiographicalDesignHCI2012}. Rather than aiming for representativeness through large user samples, autobiographical design prioritizes depth of engagement, allowing researchers to surface frictions, affordances, and breakdowns that only become visible through repeated, situated use. This framing informs the autobiographical methodology described next.

\section{Methodology}

\subsection{Research Stance}
Building on Research through Design \cite{zimmermanResearchDesignMethod2007}, we position this project as an autobiographical \cite{neustaedterAutobiographicalDesignHCI2012} material inquiry into the attention mechanism of video DiTs. We draw on our established media arts practice in experimental generative video to interrogate video DiTs from the perspective of expert practitioners. To operationalize this stance, we developed \textit{AttentionBender} as a creative probe that intervenes directly in the model’s cross-attention computation during inference. Our evidence comprises pattern-level observations from controlled variation, practice-based insights from sustained use, and novel creative artifacts produced through this workflow.

\subsection{Procedure}
\subsubsection{Implementation \& Systematic Sweep}
Our method proceeds through a cycle of implementation, generation, and interpretation. First, we developed the \textit{AttentionBender} pipeline to enable highly configurable manipulations of the attention mechanism. To understand the design space, we then employed a "grid sweep" approach, generating over 4,500 videos across orthogonal prompts (selected to vary composition and motion), seed variations, and transformation operations. Finally, we analyzed these results through comparative viewing. We constructed a custom visualization interface to read the manipulated outputs against their baselines to identify recurring visual patterns. This analytic mode follows practice-based precedents that treat systematic variation as a route to design knowledge, particularly in GenAI tools where prompt-only control is known to undermine localized intent \cite{liuDesignGuidelinesPrompt2022, zimmermanResearchDesignMethod2007}.

\subsubsection{Design Principles for the Probe}
To ensure the probe supports intuition-building rather than just random glitched outputs, the system was designed around four specific research criteria:

\begin{itemize}
    \item \textit{Legibility:} The system must clearly visualize the causality between the mathematical intervention (e.g., "Rotate all attention maps 45°") and the aesthetic result.
    \item \textit{Parametric Exploration}: It must support rapid traversal of the high-dimensional design space, moving beyond one-off trials to systemic, controlled experiments.
    \item \textit{Interpretive Lens}: The tool must help the practitioner form and revise mental models of how the intervention propagates through the network layers.
    \item \textit{Creative Affordance}: The intervention must yield novel aesthetic possibilities that are distinct from those accessible through prompt engineering alone.
\end{itemize}

\section{The Artifact}
Guided by our probe criteria (legibility, explorability, interpretive lens, creative affordance), we implemented \textit{AttentionBender} as both an intervention pipeline and a comparative viewing system.

\subsection{Bending Pipeline}
The core of the system is a modified inference loop built on top of the open-source WAN Text-to-Video model. We selected WAN for its elegant Diffusion Transformer (DiT) architecture, which processes video data as tensor embeddings passing through 30 identical transformer blocks. Each block contains self-attention, cross-attention, LayerNorm, and MLP layers.

Our intervention targets the Cross-Attention mechanism—the site where the model aligns visual latent features with text conditioning. In a standard DiT block, cross-attention calculates a relationship map between visual tokens ($Q$) and text tokens ($K$) following Eq. \ref{eq:attention}.

Normally, this attention map is treated as a flat sequence of tokens. \textit{AttentionBender} intercepts this calculation to inject a spatial transformation step. Our pipeline supports applying this intervention either pre- or post-softmax. The process follows four steps:
\begin{itemize}
    \item \textit{Reshape:} The flat visual token sequence is reshaped into its latent spatiotemporal dimensions $(Frames \times Height \times Width)$.
    \item \textit{Transform:} We apply standard 2D affine transformations (rotation, scaling, translation) or spatial filters (amplify, blur, sharpen) to each frame’s $(H \times W)$ slice. To handle "voids" created by operations like scaling down or translating, we implement configurable padding modes, defaulting to border replication to maintain visual continuity at the edges.
    \item Renormalize (Optional): Because 2D transformations can alter the sum of attention weights, we implement an optional renormalization step to ensure the attention map remains a valid probability distribution summing to 1.
    \item Flatten: The transformed map is flattened back into a sequence and passed to the Value ($V$) multiplication step.
\end{itemize}

By manipulating the spatial distribution of attention before the visual features are updated, we force the model to generate content matching the distorted attention map, highlighting the behavior of cross-attention in the generative process.

\begin{figure}[h]
  \centering
  \includegraphics[width=\linewidth]{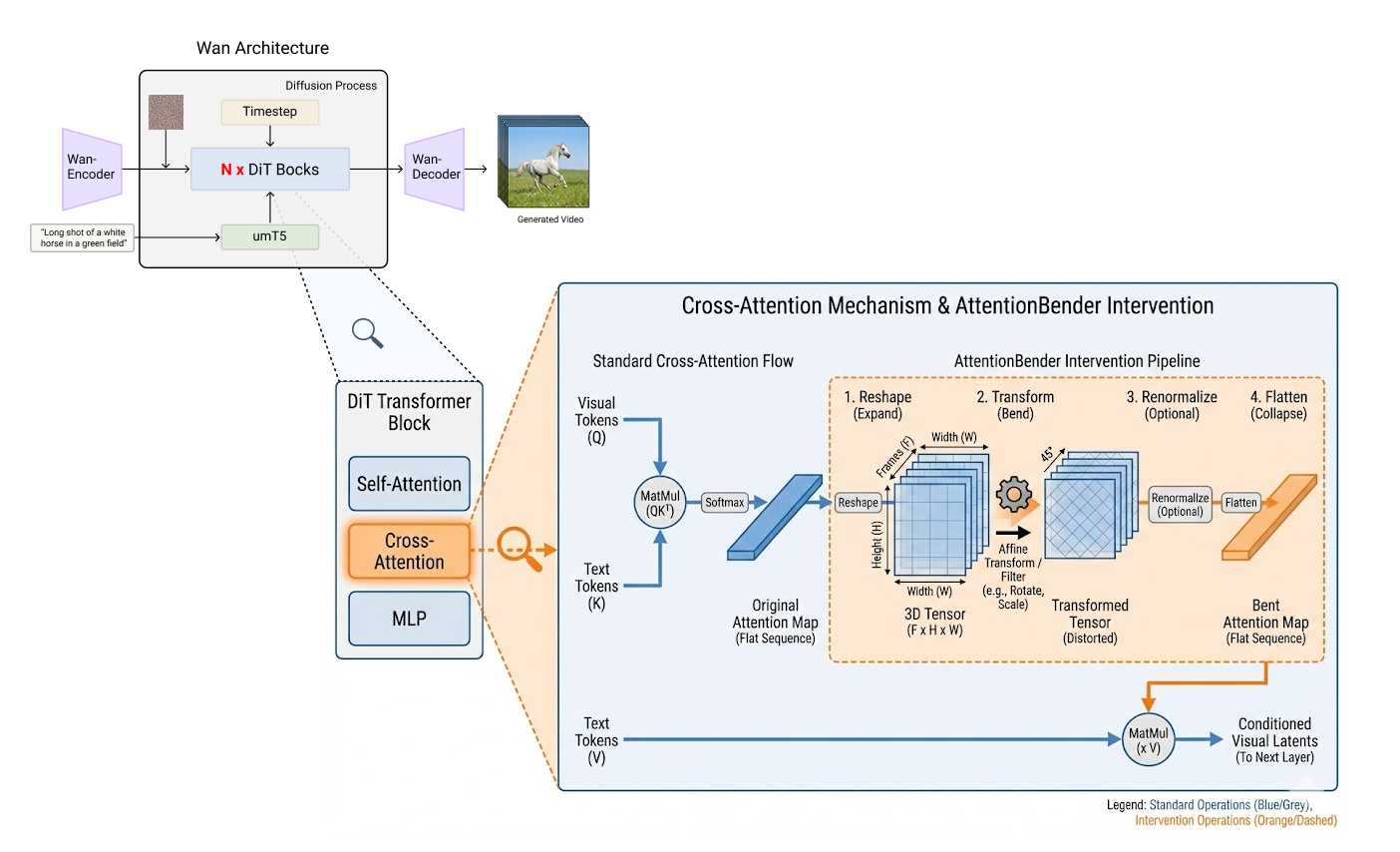}
  \caption{\textbf{Technical Diagram} of \textit{AttentionBender} pipeline. Cross-attention maps are intercepted, reshaped into 3D latent video shape $(Frames \times Height \times Width)$, transformed (translate, rotate, etc), flattened back to sequence, multiplied by Values ($V$) to complete attention process, and then continue through the standard DiT process.}
  \label{images/technical-diagram.png}
\end{figure}

\subsection{Parametric Generation Tool}
To move beyond "one-off" transforms toward a systematic understanding of model behavior, we developed a command-line generative tool controlled by a highly configurable schema. This system automates the generation of variation sets, allowing users to define generative sweeps that iterate combinatorially across five primary axes: (1) \textbf{Prompt:} The textual description driving the generation; (2) \textbf{Seed:} The random initialization noise, ensuring reproducibility, (3) \textbf{Geometric Operation:} The tool supports a suite of transform operations including scale, rotate, translate (X/Y), flip (vertical/horizontal), amplify, blur, and sharpen; (4) \textbf{Magnitude:} The user provides a range of values for a given transform and number of interval steps. For example, a range of $[-2.5, 2.5]$ with 5 steps will generate variations at $[-2.5, -1.25, 0, 1.25, 2.25]$. (5) \textbf{Targeting Masks:} To isolate specific network behaviors, the tool can selectively target specific Text Token cross-attention operations (e.g., bending only the token "Horse"), Diffusion Timesteps (intervening only in early compositional steps versus late denoising steps), and Network Layers (targeting early versus late DiT blocks).

Crucially, the tool enables storing the bent attention maps for key tokens at every timestep. We decode the flat attention maps into playable videos of dimensions $F \times H \times W$, allowing us to visualize, validate, and interpret the effect of the attention bending operation across the diffusion process. To enable organized visualization, the tool outputs a structured JSON metadata record which matches unique video filenames with the relevant transform operation configuration.

A sample configuration snippet demonstrating this schema is shown below. This specific configuration would orchestrate 192 variations for every prompt/seed combo (1 transform operation $\times$ 8 steps $\times$ 2 token targets $\times$ 3 diffusion timestep targets $\times$ 4 network layer targets):

\begin{lstlisting}[basicstyle=\ttfamily, breaklines=true]
  - operation: scale
    parameter_name: scale_factor
    range: [0.25, 4.0]   # [0.25, 0.625, 1.0, 1.375, 1.75, 2.125, 2.5, 3.0]
    steps: 8
    target_token:
      - "ALL"
      - "rose, horse, ship, actor, kiss"
    strength: 1.0
    padding_mode: "border"
    apply_to_timesteps:
      - "ALL"
      - "0-2"   # Early steps
      - "7-9"   # Late steps
    apply_to_layers:
      - "ALL"
      - "0-5"   # Early layers
      - "13-18" # Middle layers
      - "24-29" # Late layers
\end{lstlisting}

\subsection{Visualization Interface}

To interpret the high-volume outputs characteristic of systematic generative sweeps, we developed a web-based visualization interface designed to support comparative analysis across this large parametric space (Fig.~\ref{fig:full_grid}). The interface organizes results into a navigable grid where each row represents a specific transform operation and magnitude, and each column represents a unique Prompt/Seed combination. Pinned to the top of the grid are the baseline results: standard generations produced without any attention bending. This layout encourages two distinct modes of reading. Scrolling horizontally allows us to observe the consistency of a transform’s effect across different prompts and seeds, helping us distinguish between consistent structural patterns and idiosyncratic variations unique to specific random seeds. Scrolling vertically reveals how modulating the transform value impacts the generation, providing an intuition for how the network responds to increasing distortion—for example, observing the progressive deformation of a subject as a scaling factor increases from 0.25x to 4.0x.

\begin{figure}[h]
  \centering
  \includegraphics[width=\linewidth]{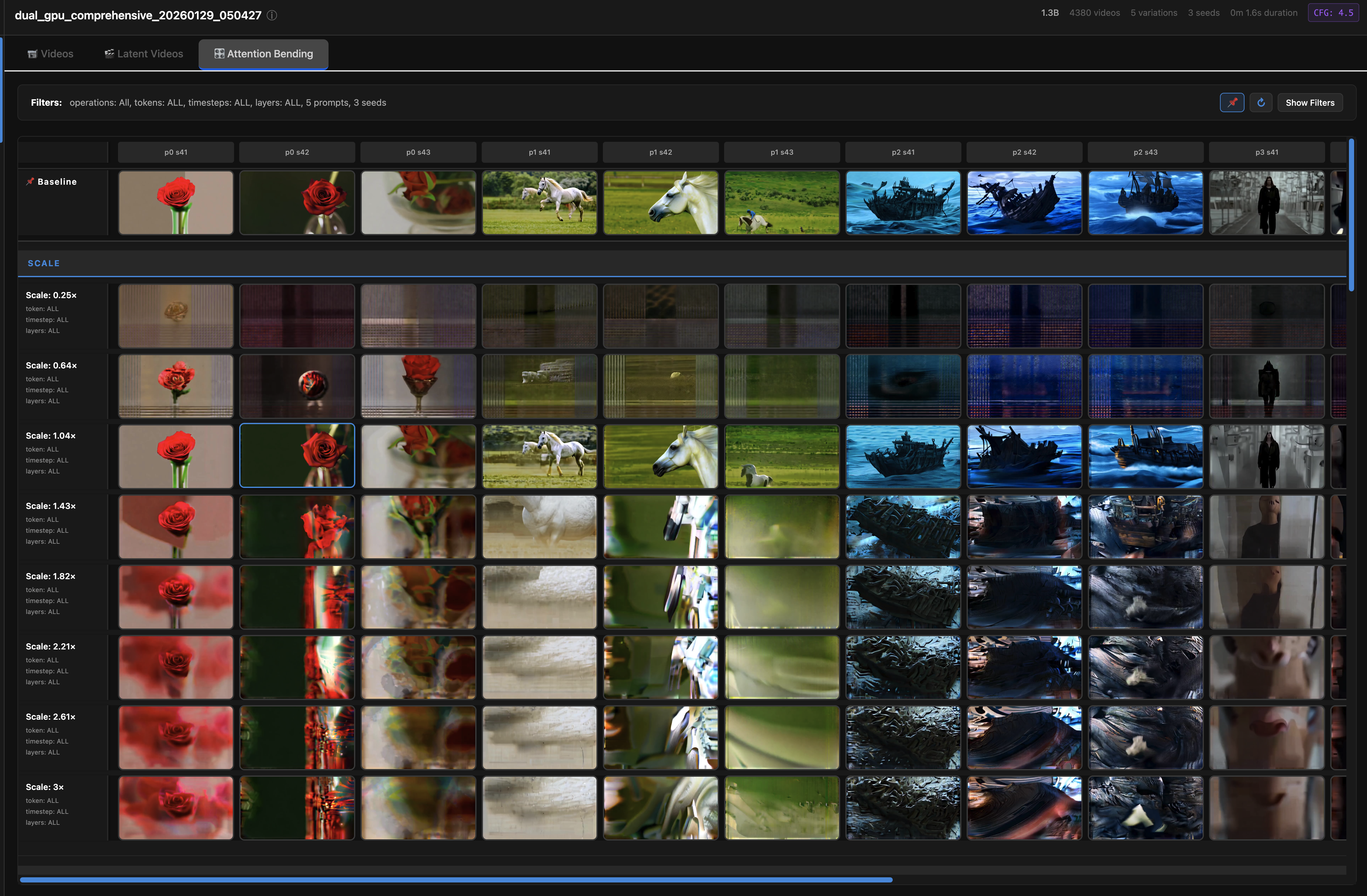}
  \caption{\textbf{The Comparative Visualization Interface.} A grid-based tool for analyzing the generative sweep. Columns organize outputs by prompt and seed, while rows organize them by intervention type and magnitude relative to a pinned baseline.}
  \label{fig:full_grid}
\end{figure}

Navigating this extensive dataset requires precise isolation of variables. A robust Filter Menu (Fig.~\ref{fig:filters}) enables targeted investigation, allowing us to toggle visibility based on specific Operations, Token Targets, Timestep Ranges, and Layer Ranges. This functionality is essential for comparative study; for instance, we can isolate experiments targeting the single token "Horse" versus those targeting "ALL" tokens to discern the distinct impact of localized versus global attention manipulation.

\begin{figure}[h]
  \centering
  \includegraphics[width=\linewidth]{images/Filters.png}
  \caption{\textbf{Filter Menu.} Enables targeted navigation of the dataset by isolating specific operations, tokens, timesteps, and layer ranges to facilitate comparative analysis.}
  \label{fig:filters}
\end{figure}

Clicking on any generation opens a Drill-Down Inspection view (Fig.~\ref{fig:video_details}). This modal presents the generated video side-by-side with its corresponding baseline, enabling direct visual comparison of the intervention's impact. Below the video player, the interface visualizes the bent attention maps for the targeted tokens across all diffusion timesteps. By exposing the internal attention dynamics alongside the aesthetic output, this view connects the final video directly to the site of intervention, supporting an interpretive reading of how spatial distortions in the attention mechanism propagate through the diffusion process.


This design supports rapid skimming and close inspection by making it possible to trace causal relationships between intervention and outcome (legibility), while also enabling the discovery of outputs that diverge most strongly from the baseline. In this way, the interface supports both analytic sensemaking and creative exploration within a large, high-dimensional design space.

\section{Findings}

\subsection{Experimental Setup}
\label{sec:exp-setup}
To map the material boundaries of the Video Diffusion Transformer, we conducted a systematic generative sweep across the parameter space defined by our tool. Generations were performed using the WAN 2.1 (1.3B-parameter) video model at a resolution of $368 \times 640$ px and a duration of 25 frames, utilizing 10 diffusion steps with a classifier-free-guidance (CFG) of 6.5. The experiment was distributed across two NVIDIA RTX A6000 GPUs. 

We selected 5 distinct prompts (enumerated in Appendix ~\ref{sec:appendix_schema}) representing common video generation use cases (e.g., "A horse galloping," "A cinematic close-up"). For each prompt, we generated variations across 3 random seeds to distinguish consistent mechanical behaviors from random seed-based artifacts. These parameters were selected to sample representative regions of the design space rather than to exhaustively enumerate all possible behaviors. The sweep covered a matrix of 9 transformation operations (scale, rotate, translate X/Y, flip H/V, blur, sharpen, amplify), applied across varying magnitudes, target tokens, timestep ranges, and layer depths. In total, this produced a corpus of 4,560 videos. Transforms were applied to attention maps post-softmax operation and were re-normalized before multiplication with the \textit{value} tensor. The full configuration schema is detailed in Appendix~\ref{sec:appendix_schema}.

Below, we present a taxonomy of recurring behaviors observed across this corpus. We then translate these observations into practice, showing how intuition built through the sweep can guide the deliberate production of creative artifacts. Due to space, we show representative slices; the full sweep grids are available online  \url{https://attention-bender.netlify.app/}.

\subsection{Experimental Results}

\subsubsection{Linearity and its Limits: 2D Transforms}
\label{sec:2d-transforms}
A primary motivation for this probe was to test whether geometric transforms applied to cross-attention maps can meaningfully steer composition in the output. Across translation, scaling, rotation, and flipping, we found a consistent directional effect on composition across prompts and seeds, but rarely a clean object-like transform. Even at mild magnitudes, deformation is common. What remains stable is the predictable direction of the compositional change.

\begin{figure}[!ht]
  \centering
  \includegraphics[width=\linewidth]{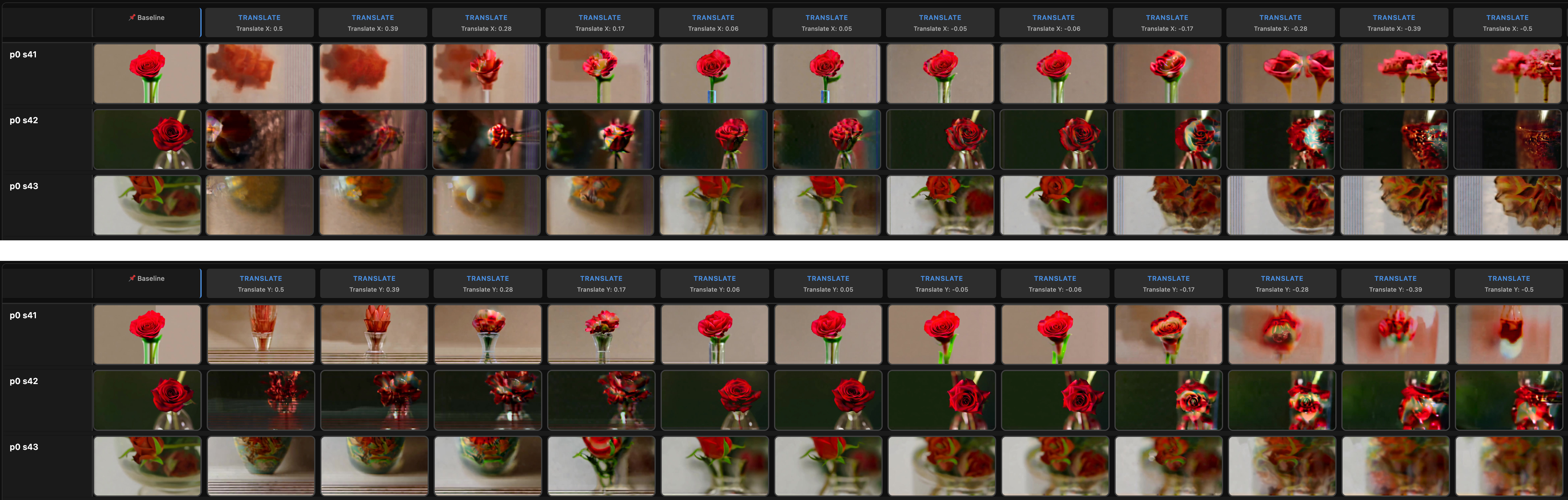}
  \caption{\textbf{Translation} consistently shifts the position of the subject across the frame, but representation degrades quickly and unevenly as offsets grow.} 
  \label{fig:translate}
\end{figure}

\textbf{Translation shifts composition, but coherence degrades unevenly.}
Translating cross-attention maps across all tokens produces a consistent shift in the scene’s focal region (Fig.~\ref{fig:translate}). As translate\_x moves from $-0.5$ to $0.5$, the salient content shifts left-to-right; translate\_y produces a corresponding top-to-bottom shift. However, this does not behave like moving a cut-out object in an editor. As we deviate from the baseline, subject coherence often collapses: subjects smear, fragment, or dissolve and re-form. Occasionally the shift reads as a meaningful reframing (e.g., $p0\,s41$, translate\_x $=0.5$ yields a flower cleanly positioned on the right), but many neighboring settings produce distortion without a stable “moved subject.”

\begin{figure}[!h]
  \centering
  \includegraphics[width=\linewidth]{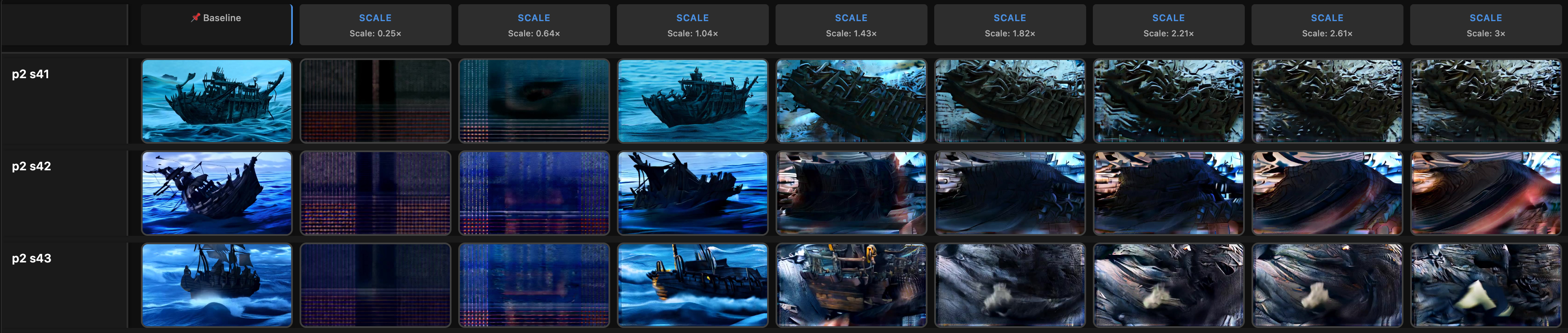}
  \caption{\textbf{Scaling} changes the spatial footprint of the prompt’s focal content. Small changes can produce subtle, legible changes in subject size, while larger scale factors rapidly reduce representational coherence.} 
  \label{fig:scale}
\end{figure}

\begin{figure}[!h]
  \centering
  \includegraphics[width=\linewidth]{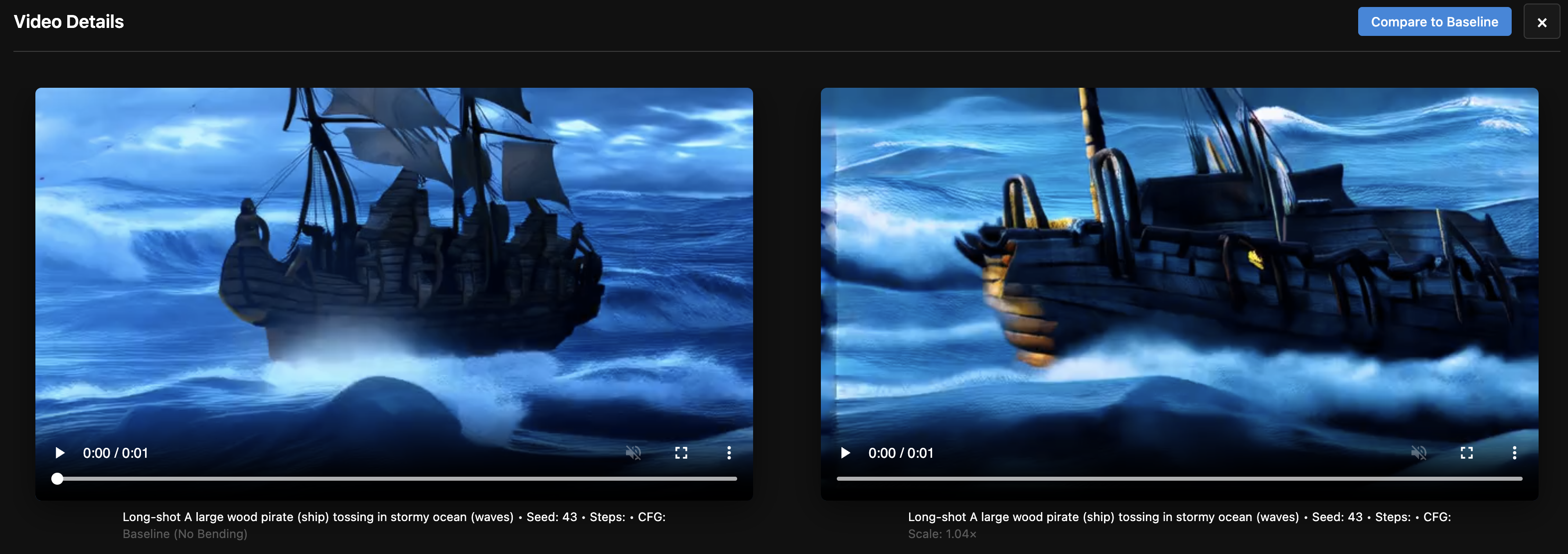}
  \caption{\textbf{A small scale increase} (1.04$\times$) shows a subtle but legible difference in subject size relative to the baseline, illustrating the narrow range where scaling yields a meaningful edit.} 
  \label{fig:scale-detail}
\end{figure}

\textbf{Scaling expands or compresses the scene, with a narrow window of coherent change.}
Scaling the attention maps changes the spatial footprint of where the prompt concentrates the subject within the frame (Fig.~\ref{fig:scale}). In the ship example, a small scale increase (e.g., $1.04\times$) produces a subtle but legible change in subject size relative to the baseline (Fig.~\ref{fig:scale-detail}). As the scale factor increases further, the output’s content becomes progressively less coherent, often turning into heavy deformation rather than a larger-but-stable object.

\begin{figure}[!h]
  \centering
  \includegraphics[width=\linewidth]{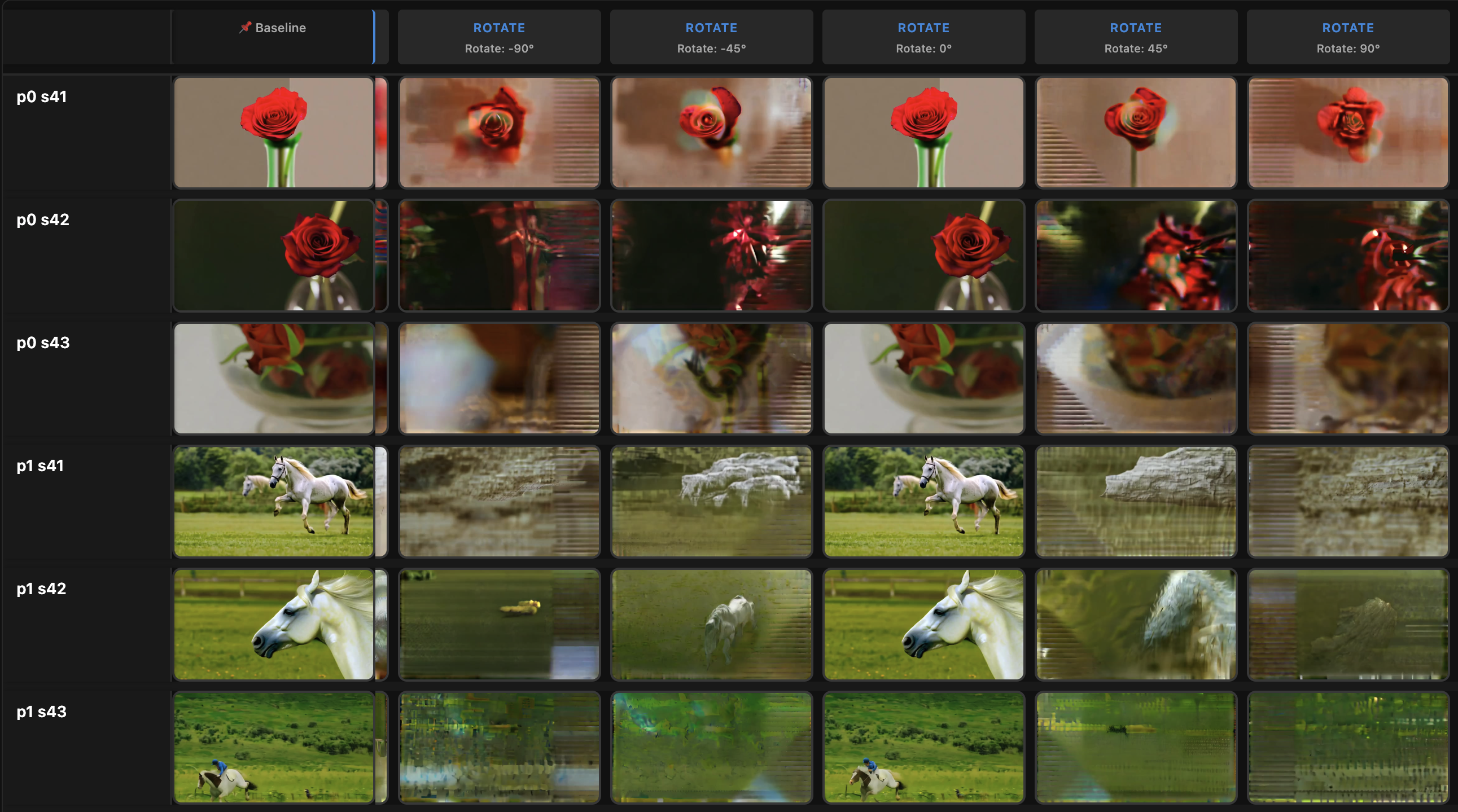}
  \caption{\textbf{Rotate} increasingly degrades coherence with greater angle degree as opposed to stable object rotation; Operation more destructive for asymmetrical compositions than radially symmetric videos.}
  \label{fig:rotate}
\end{figure}

\textbf{Rotation increases deformation with angle.}
As the rotation angle grows, outputs lose representational coherence rather than producing a stable rotated scene (Fig.~\ref{fig:rotate}). Symmetric subjects (e.g., the flower in $p0\,s41$) sometimes retain partial coherence, while asymmetrical scenes tend to break down more sharply.

\begin{figure}[h]
  \centering
  \includegraphics[width=\linewidth]{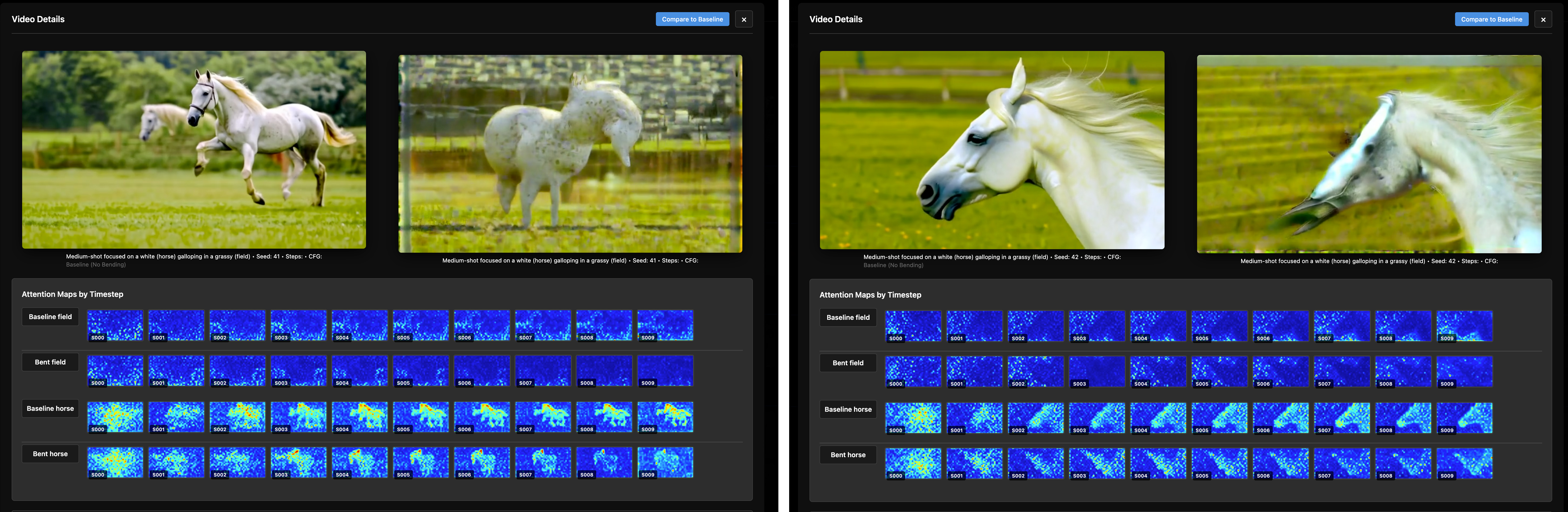}
  \caption{\textbf{Flipped} cross-attention maps do not behave like a perfect mirror in the output. Notably, while horizontal flips preserve a broadly mirrored layout with distortion, vertical flips fail to produce a stable upside-down subject, instead deforming and re-arranging recognizable features.} 
  \label{fig:flip}
\end{figure}

\textbf{Flip operations manipulate spatial layout, but not canonical orientation.}
The flip transformation makes the gap between attention-map geometry and semantic structure especially visible (Fig.~\ref{fig:flip}). When we flip attention maps horizontally, the resulting outputs often show a broadly mirrored layout, but with major representational distortion. However, when we flip vertically, we do not obtain a stable upside-down subject. Instead, recognizable features persist but are re-arranged in unexpected ways: in our horse example, features like an eye or mane may remain identifiable, while other parts deform dramatically, like the nose region becoming hoof-like.

\subsubsection{Texture Control: Blur, Sharpen, Amplify}
Beyond geometric transforms, we observed consistent patterns in filter-style operations (blur, sharpen, amplify) and the level of detail present in the outputs. Operations that accentuate attention distributions and contrast tended to coincide with sharper textures and more visible detail, while operations that smooth the maps tended to coincide with flatter, less articulated surfaces.

\begin{figure}[h]
  \centering
  \includegraphics[width=\linewidth]{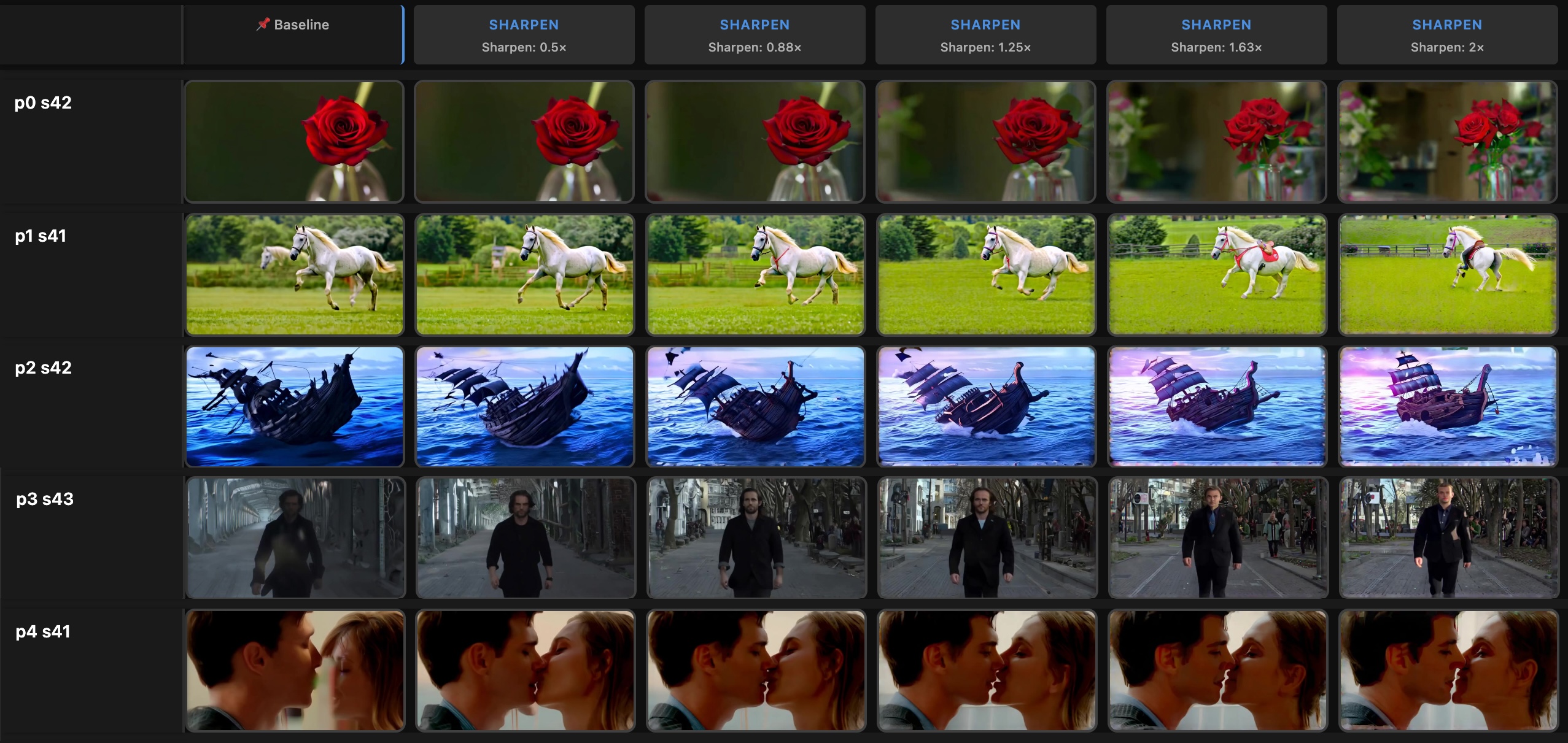}
  \caption{\textbf{Sharpen} increases perceived detail and definition across subjects and backgrounds. Compared to baseline, outputs show more articulated textures (e.g., petals, sails, facial features) and often increased background structure.}
  \label{fig:sharpen}
\end{figure}

\textbf{Sharpen increases perceived detail, often affecting background as well as subject.}
Sharpening the attention maps reliably produced outputs with more defined content and richer surface texture across prompts and seeds (Fig.~\ref{fig:sharpen}). In these examples, rose petals, ship sails, and facial features appear more articulated relative to baseline. The effect is not confined to the primary subject: background regions often gain additional structure (e.g., added urban detail behind the figure in $p3\,s43$). In the attention-map visualizations, sharpened maps show stronger contrast and finer grain than baseline, which coincides with the increased “crispness” observed in the resulting videos.

\begin{figure}[h]
  \centering
  \includegraphics[width=\linewidth]{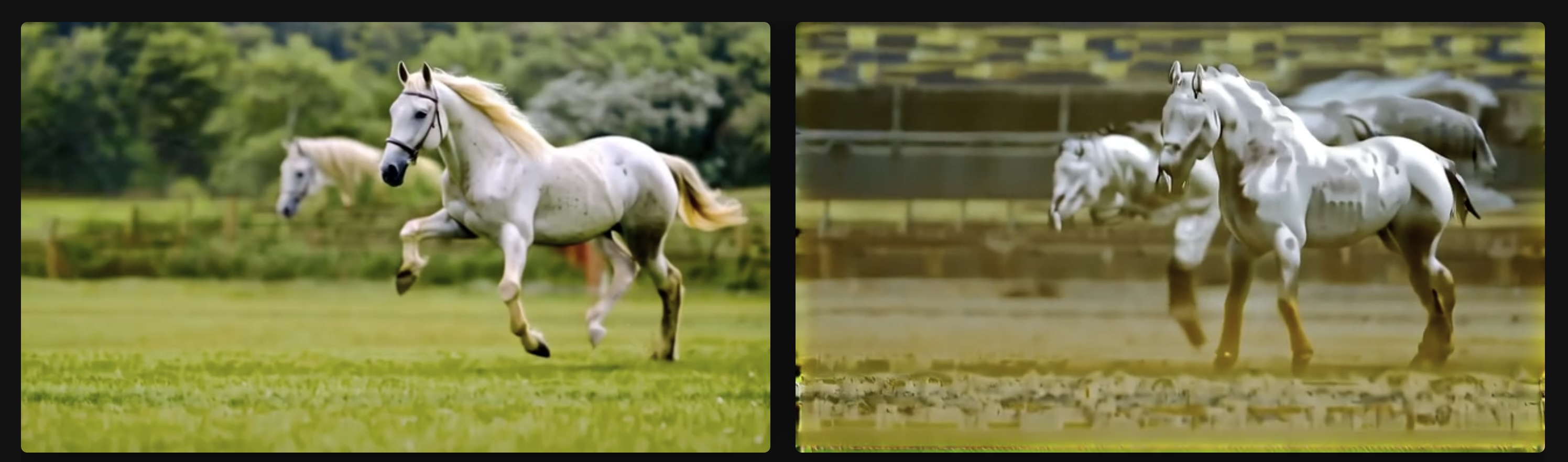}
  \caption{\textbf{Blur} reduces fine-grained texture without behaving like a standard image blur. Composition often remains similar to baseline, while surfaces become smoother and less differentiated (e.g., foliage, hair, facial features).}
  \label{fig:blur}
\end{figure}

\textbf{Blur reduces fine detail without behaving like a conventional video blur.}
Blurring the attention maps did not simply “blur” the image in a post-processing sense. Instead, it tended to reduce textural complexity while leaving overall composition largely intact (Fig.~\ref{fig:blur}). Elements such as fur, foliage, and facial features often become smoother and less differentiated, even when composition and object boundaries remain similar to baseline. This corresponds to attention maps that appear more smoothed, with weaker contrast between higher and lower weighted regions.

\begin{figure}[h]
  \centering
  \includegraphics[width=\linewidth]{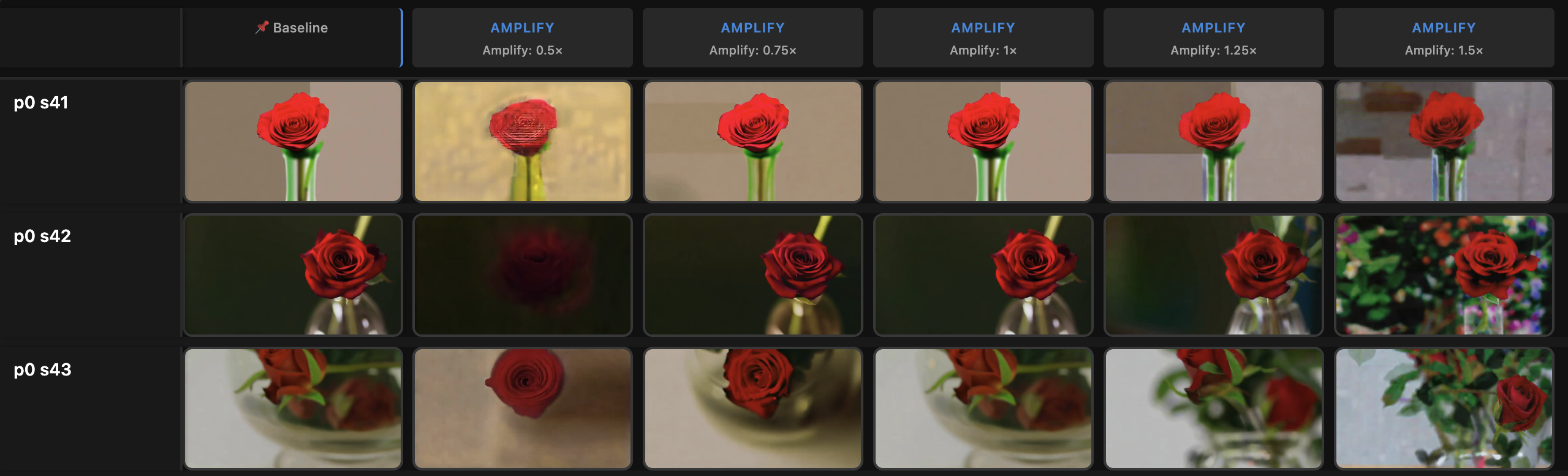}
  \caption{\textbf{Amplify} increases subject detail and visible representation, with extreme values affecting the spatiotemporal domain.}
  \label{fig:amplify}
\end{figure}

\textbf{Amplify acts as a coarse intensity control.}
Amplify showed a similar directional effect (Fig.~\ref{fig:amplify}). Values below $1.0$ often dampened surface detail, while values above $1.0$ tended to intensify texture and contrast. At more extreme values (e.g., $2.0\times$), we occasionally observed changes in subject motion  (faster or more erratic movement), suggesting that amplification can influence temporal behavior as well as surface detail.

\subsubsection{System Resilience: Token, Timestep, \& Layer Targets}
A striking pattern across the sweep was the model’s tendency to remain visually stable under many interventions. Across token-, timestep-, and layer-targeted conditions, bending cross-attention often produced weaker or non-proportional changes in the output, even when the attention maps were heavily transformed.

\begin{figure}[!ht]
  \centering
  \includegraphics[width=\linewidth]{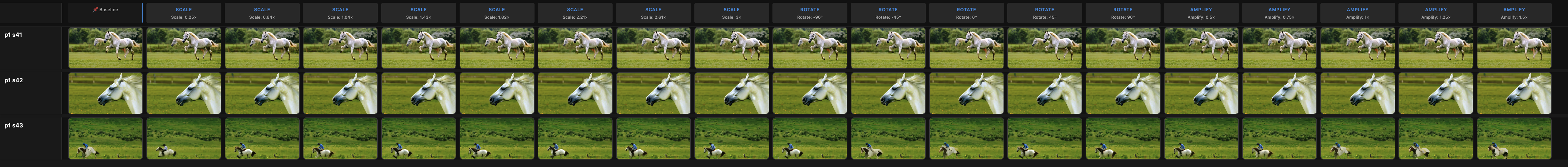}
  \caption{\textbf{Single-token targeting.} Across multiple operations, bending cross-attention for a single subject token often produces outputs close to baseline, compared to \texttt{ALL}-token interventions.}
  \label{fig:single-token}
\end{figure}

\textbf{Single-token interventions often produce minimal visible change.}
Targeting individual tokens frequently produced minimal visible change compared to targeting \texttt{ALL} tokens (Fig.~\ref{fig:single-token}). Even when the attention map for the primary noun was heavily transformed, outputs often remained close to baseline across operations.

\textbf{Early timesteps are more sensitive than late timesteps.}
We observed a clear timestep dependence consistent with diffusion’s coarse-to-fine dynamics. Interventions applied during early timesteps (0--2) more often altered layout and coherence, while interventions during late timesteps (7--9) were frequently muted or limited to minor artifacts.

\begin{figure}[!h]
  \centering
  \includegraphics[width=0.5\textwidth]{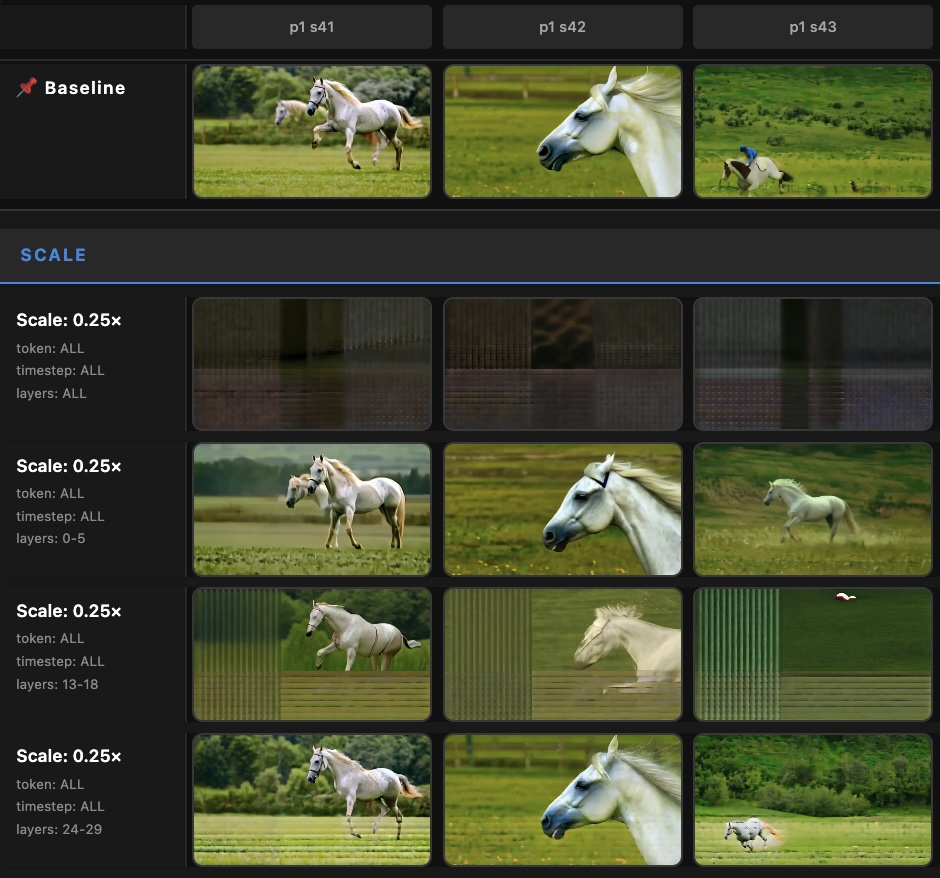}
  \caption{\textbf{DiT-layer targeting.} Outputs remain comparatively coherent when modulating subsets of the DiT layers; perturbation of the middle layers has largest effect.}
  \label{fig:layer-target}
\end{figure}

\textbf{Layer targeting reduces deformation, with middle blocks most influential.}
A similar pattern appears when restricting interventions to subsets of transformer blocks. Targeting only early (0-5), middle (13-18), or late layers (24-29) generally produced significantly weaker deformations than bending all layers (Fig.~\ref{fig:layer-target}), with many outputs retaining recognizable structure. Intervening in layers 13-18 tended to produce the strongest visible changes relative to comparable interventions in early or late blocks. 

\subsection{Emergent Aesthetic Patterns and Opportunities} 

Beyond the study of control, our autobiographical engagement with the probe revealed aesthetic behaviors that sit noticeably outside the model’s typical prompt-based aesthetics. Across these artifacts, the key point is not that attention bending “makes interesting glitches,” but that the glitches are \emph{structured} in a legible way: they preserve relationships to the intervention (direction, symmetry, and shifts in texture) that echo the compositional and textural dynamics described above. As our intuition for the parameter space improves — what different transforms tend to do, where coherence collapses, and how effects interact across magnitude — we can intentionally steer the system toward specific aesthetic spaces rather than treating breakdown as a random failure mode.

\subsubsection{Exploding Poetic Pixels}
When pushed toward extreme magnitudes, 2D transform operations frequently drive the output into dramatic non-representational color fields. Saturated textures, smeared motion, and pixel-burst artifacts foreground the generative material in its own right. These results resonate with traditions of abstract expressionism and glitch aesthetics, and for us they most strongly evoke the transcendent, computational films of \textit{Jordan Belson} \cite{belsonSamadhi1967}. 

\begin{figure}[h]
  \centering
  \includegraphics[width=\linewidth]{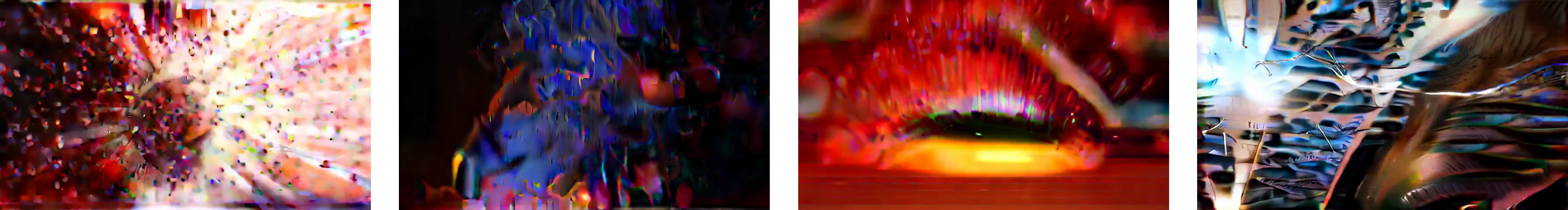}
  \caption{\textbf{Exploding Poetic Pixels.} Extreme geometric interventions push outputs into saturated abstraction while preserving a legible relationship to the transform: flip introduces bilateral symmetry; translate$_y$ smears mass directionally; extreme scale inflates ship-and-wave textures into a volatile abstract composition.}
  \label{fig:abstract-composite}
\end{figure}

Importantly, these abstractions are not arbitrary. They retain a legible relationship to the intervention. In Fig.~\ref{fig:abstract-composite}, the leftmost image is generated from a horizontal flip, producing a deformation with recognizable bilateral symmetry. The middle two images are translate operations along the vertical axis, smearing the scene’s mass downward or upward in a manner consistent with the directional compositional shifts observed in Sec.~\ref{sec:2d-transforms}. The rightmost image comes from an extreme scale operation on the sailing ship prompt: while the output becomes largely non-representational, it retains faint traces of wave- and hull-like contours, linking the abstraction back to the prompt and to the material effect of the transform.

\subsubsection{Extreme and Ambiguous Representations}
At medium magnitudes, where coherence has not fully collapsed, attention bending can produce representations that are unstable but still readable: faces and bodies that drift into ambiguity, excess, or unsettling smoothness. In this context, the filter operations are especially useful because they let us tune the balance between representation and deformation.

\begin{figure}[h]
  \centering
  \includegraphics[width=\linewidth]{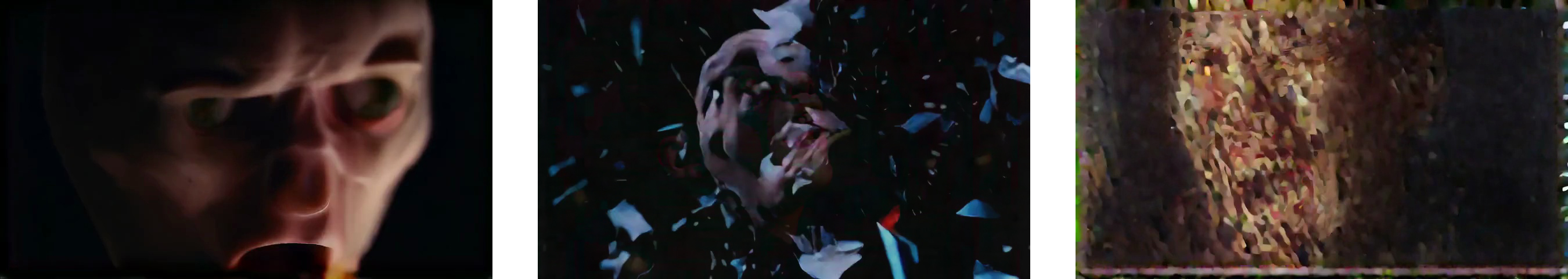}
  \caption{\textbf{Extreme and Ambiguous Representations.} Filter operations used to balance legibility and abstraction. Left: blur smooths facial structure into the uncanny. Center: amplify exaggerates feature intensity. Right: sharpen dissolves the face into dense, brittle texture.}
  \label{fig:representational-composite}
\end{figure}

Fig.~\ref{fig:representational-composite} shows three variations that use bending filters to navigate that boundary. The left panel applies blur to cross-attention, yielding an eerily smoothed face that largely preserves composition while shedding flesh texture. The center panel uses amplify to push toward a more explosive, high-energy rendering where features flare. The right panel uses sharpen to drive the face toward a brittle textural dissolution while retaining a ghost of the original composition.

\begin{figure}[h]
  \centering
  \includegraphics[width=\linewidth]{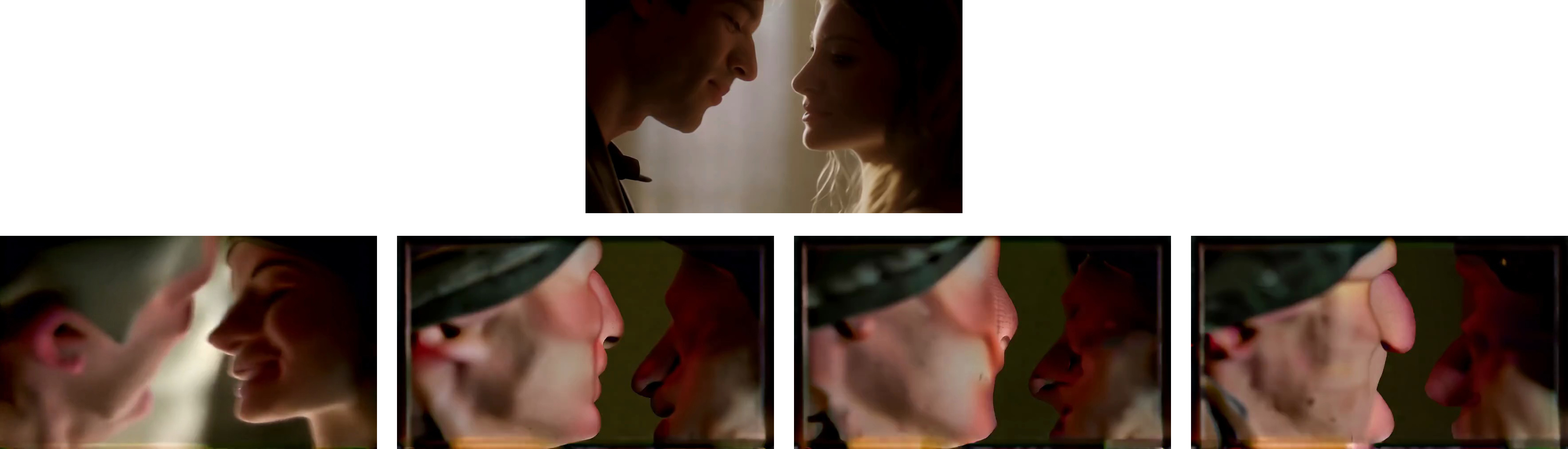}
  \caption{\textbf{Parametric axis as creative structure.} Increasing blur strength modulates a generic cinematic kiss into a progressive flattening and destabilization of a familiar trope.}
  \label{fig:kiss-blur}
\end{figure}

Finally, Fig.~\ref{fig:kiss-blur} illustrates how parametric control can become an expressive creative affordance. Starting from a baseline prompt of a romantic kiss (generic, cinematic, and recognizably hetero-normative), we apply the blur filter at increasing strengths. As blur increases, the lovers flatten and simplify, producing a progression that is both unsettling and visually compelling. The sequence makes it possible to \emph{read} what resists deformation (iconic kiss pose, subtle gender markers) and what dissolves first (facial detail, unique identities, and fine texture), turning a familiar trope into a controlled drift toward provocative ambiguity.

\section{Discussion}
\subsection{The Materiality of the Model: Building New Intuitions}

\subsubsection{Cross-Attention: Syntax vs. Semantics}

A primary insight from this study is the functional limit of cross-attention in the generative pipeline. Across our sweep, we observed that cross-attention behaves less like a geometry engine and more like a \emph{spatial distributor}: it nudges where visual energy pools, but it does not strictly dictate the geometric reality of the object.

This tension is most visible in the failure of the vertical flip operation (Sec.~\ref{sec:2d-transforms}). While we successfully inverted the attention map, the model did not generate an upside down world. Flower petals stayed above the stem, horse heads remained above feet, and the sky remained above ground. Rather than object inversion, the generation reorganizes into distorted but recognizable fragments that gravitate back toward familiar orientations. This suggests a division of labor within the architecture: Cross-attention provides the \emph{content syntax} (the rough spatial distribution), while the subsequent value vectors and feed-forward layers likely enforce the \emph{content semantics} (the canonical appearance and physics of the object).

For us artist-practitioners, this is a crucial intuition. \textit{AttentionBender} makes this theoretical distinction tangible, allowing the artist to feel the friction between the spatial suggestion of the prompt and the semantic rigidity of the model's training data.

\subsubsection{Material Flexibility and Elasticity}
We initially considered that the attention mechanism might be brittle: that a slightly disturbed map would immediately collapse the generation into noise. Instead, we found the system exhibited a surprising resilience: a property we frame as  \emph{material flexibility}. Small manipulations were absorbed without effect, and even more global distortions often failed to break the subject's coherence entirely.

This observation suggests a redundancy within the system. For example our token-targeting experiment suggests that the "impact" of a specific prompt-word is not localized solely in the cross-attention token we targeted. Plausibly, semantic cues are already encoded by the text encoder and distributed through self-attention and MLP layers before the cross-attention operation occurs. This creates a distributed flexibility that allows the model to resist our interventions.

This resilience to deformation offers us a creative affordance. For example, in our "Kiss" study (Fig.~\ref{fig:kiss-blur}), we were able to apply filters of increasing strength without the outputs collapsing into noise; rather, the model exhibited a "self-healing" tendency, using its internal priors to reconstruct a coherent image within a distorted process until the breaking point is reached. This resilience turns the generative black-box into an elastic, flexible medium with wide-ranging creative outputs beyond its rigid defaults. 

The \textit{AttentionBender} workflow takes advantage of this quality, allowing us to push against the model's normative defaults. Our creative examples show that by modulating the network bending strength, we can exploit the boundary between the model's attempt to "fix" the image and its ultimate collapse into abstraction for novel artistic purposes.

\subsubsection{From Equation to Intuition}
Ultimately, this inquiry demonstrates the value of Network Bending \cite{broadNetworkBendingExpressive2021}style probing tools for building practice-based intuition. While the mathematical definition of attention, $Attention(Q,K,V)$, is well-documented, the \emph{behavioral reality} of that equation is inaccessible through standard interfaces or theory alone.

By systematically intervening in the generation process, \textit{AttentionBender} transforms the abstract equation into a tangible material property. It allows us to isolate the influence of the query-key relationship ($QK^T$) from the semantic payload of the values ($V$). This validates research-through-design as a bridge between technical explainability and creative practice: by building the tool to break the model, we gained a more grounded, hands-on intuition of how the model works.

\subsection{Designing for Material Intuition}

In our autobiographical practice, we found that building intuition for a generative model required more than just observing inputs and outputs; it required an interface that bridges the gap between high-level parameter navigation and low-level mechanism inspection.

What began as a command-line script evolved into a dual-view interface because we found that neither view was sufficient on its own. The high-level grid (the "Macro" view) was necessary to identify stable patterns like the \emph{syntax/semantics} split, distinguishing consistent model behaviors from random seed artifacts. However, interpreting \emph{why} a pattern occurred (such as the the increased detail from sharpen operations) required drilling down into the individual attention maps (the "Micro" view). For instance, regarding the \textit{sharpen} operation, the macro view showed us \emph{that} the video became detailed, but the micro view showed us \emph{how}: visualizing the attention map at each step of the diffusion process revealed that attention maps with greater texture and contrast corresponded with the model filling in more granular content in the output.

Reflecting on this workflow, we propose that the set of design principles used for \textit{AttentionBender} may be relevant for future XAIxArts tools \cite{bryan-kinnsExploringXAIArts2023}. Our experience suggests that interfaces should couple \textbf{high-parametric explorability} (to map the space of plausible outputs) with \textbf{mechanism legibility} (to expose the internal state). By visualizing the intermediate generative steps and not just the final pixels, tools can transform the "black box" into a more legible, interpretable process, allowing artists to develop their creative instinct with rigor.

\subsection{Toward Material Intimacy} 
As demonstrated in \ref{sec:artist-practice-video}, generative video tools keep practitioners at arm's length from the actual model mechanics, mediating the relationship through language and web interfaces \cite{cetinicUnderstandingCreatingArt2022}.

\textit{AttentionBender} argues for a different relationship: \textit{material intimacy}. To understand generative video as a medium, we must be able to reach inside the machinery, to touch the weights, tensors, attention maps and everything else which make up a model. 

This aligns our work with the lineage of experimental structural film, where artists like Stan Brakhage physically interfered with the celluloid strip to reveal the medium's photochemical reality \cite{brakhageMothlight1963, sitneyVisionaryFilmAmerican2002}, and media artists like Nam June Paik modulated the cathode-ray signal to reveal the electronic beam beneath the broadcast image \cite{paikMagnetTV1965}. By adopting network bending to directly manipulate the cross-attention mechanism, we aim to establish a similar tactile relationship with the neural network

Beyond a materialist critique, our interventions also engage the model’s ideological defaults (what it repeatedly treats as normal, stable, and physically plausible) by pressure-testing them under controlled deformation. Our findings demonstrate the persistence of familiar representations under stress, from the orientation of the horse (Fig.~\ref{fig:flip}) to the iconic silhouette of the kiss (Fig.~\ref{fig:kiss-blur}). Paying attention to what artifacts disappear and what persist during our interventions becomes diagnostic, exposing the canonical figures that the model holds onto under pressure. This resonates with Glitch Art practices \cite{menkmanGlitchMomentum2011}, where "failure"  is used as a method to expose the underlying technical and cultural assumptions of the technology.

We contend that future creativity support tools should aim for this level of openness. We want to see designs that move from \textit{prompt engineering} (a mode of indirect request) to a mode of \textit{material engagement}, where the artist has a hands-on relationship with the internal states of the system. Instead of being passive users, limited by a prompt-box or even the model's inherent representations, we aim for tools that develop a sense of agency, ownership and radical possibility.

\subsection{Limitations and Future Work}
This work is an autobiographical Research through Design probe \cite{zimmermanResearchDesignMethod2007}. Our findings are grounded in our individual expert practice and in one specific model configuration (Wan Video, 1.3B). As such, we do not present these observations as general laws of video diffusion transformers. Instead, we offer them as a mapped set of behaviors from a concrete intervention into a particular system, intended to support intuition-building and to motivate further study.

A second limitation is architectural. Many of the behaviors we surfaced point to entanglement and redundancy that actively resist localized control. In practice, this is both a constraint and a clue for future research opportunities. It suggests that some “edit-like” goals may be structurally misaligned with cross-attention bending alone, and that creative openness may require new or augmented intervention modes, either within existing architectures or through models designed with artist-facing interpretability in mind.

Several extensions follow directly from this. First, we can test generality by widening the design space: larger Wan models (e.g., 14B), alternative video diffusion systems (e.g., HunYuan \cite{kongHunyuanVideoSystematicFramework2025}), and varied generation settings (steps, CFG, resolution, prompt families). Second, our current bending operations apply uniformly per-frame. A natural next step for video DiT manipulation is temporal bending: applying transforms that vary over frames rather than only across $(H \times W)$ to explore motion as a first-class affordance. Third, guided by our syntax/semantics split, future work should extend beyond the attention map itself to other sites that may carry  more “what” than “where,” including the Value ($V$) vectors, MLP pathways, and related conditioning routes, in order to build deeper intuitions for the complete diffusion transformer process.

Finally, we see a clear opportunity to expand impact beyond a personal probe. We are releasing \textit{AttentionBender} as open-source code at <REDACTED-FOR-ANONYMITY>, alongside our representative results at \url{https://attention-bender.netlify.app/}, so other artists and researchers can reproduce, challenge, and extend these observations. By moving from an autobiographical instrument to a community resource, we hope to encourage shared vocabulary and comparative practice around how artists can reclaim agency within otherwise opaque generative video systems.

\section{Conclusion}
This paper presented \textit{AttentionBender}, an inference-time network bending pipeline for video diffusion transformers that applies spatial transforms to cross-attention maps. To make this intervention usable as a creative probe, we paired the pipeline with a systematic sweep workflow and a comparative visualization interface that supports high-level filtering alongside drill-down inspection of outputs and attention dynamics.

Using this workflow, we report pattern-level observations from a 4,560-video corpus. While \textit{AttentionBender} did not demonstrate clean linear edits of subjects, we observed meaningful directional change, from redistributing visual focus in a scene to modulating scene texture. Using the sweep to read these patterns built practical intuition about the generative process, which we then leveraged to create novel artifacts that push beyond the model’s normative domain.

Together, these contributions suggest a reusable method for artist-facing interpretability: intervene on a legible internal mechanism, map its behavior through systematic variation, and make the space readable through comparative tools. The goal is not surgical control, but \textit{material intimacy} — an approach that can extend across architectures and modalities to support intuition-building and expressive departures beyond prompt-based workflows.

\begin{acks}
Adam Cole’s research is supported by the UKRI Techné Studentship, AHRC Grant reference number AH/R01275X/1.
\end{acks}

\bibliographystyle{ACM-Reference-Format}
\bibliography{PhD-bibtex}

\appendix

\section{Experiment Configuration Schema}
\label{sec:appendix_schema}

The following configuration schema illustrates the structure used to orchestrate the systematic sweep described in Section \ref{sec:exp-setup}. This YAML structure defines the parameter ranges, step counts, and target layers for the \textit{AttentionBender} inference pipeline.

\begin{lstlisting}[basicstyle=\ttfamily\small, breaklines=true, frame=single, caption={AttentionBender Sweep Configuration}]
attention_bending_settings:
  apply_before_softmax: false
  apply_to_output: true
  configs: []
  device: cuda
  enabled: true
attention_bending_variations:
  enabled: true
  generate_baseline: true
  renormalize: false
  operations:
  - operation: scale
    parameter_name: scale_x
    range: [0.25, 3.0]
    steps: 8
    target_token:
    - "ALL"
    - "rose, horse, ship, actor, kiss"
    apply_to_timesteps:
    - "0-2"
    - "7-9"
    - "ALL"
    apply_to_layers:
    - "0-5"
    - "13-18"
    - "24-29"
    - "ALL"
    strength: 1.0
    padding_mode: border
  - operation: rotate
    parameter_name: angle
    range: [-90.0, 90.0]
    steps: 5
    target_token:
    - "ALL"
    - "rose, horse, ship, actor, kiss"
    strength: 1.0
    padding_mode: border
  - operation: translate
    parameter_name: translate_x
    range: [-0.05, 0.5]
    steps: 6
    strength: 1.0
    padding_mode: border
  - operation: translate
    parameter_name: translate_y
    range: [-0.05, 0.5]
    steps: 6
    strength: 1.0
    padding_mode: border
  - operation: flip
    parameter_name: flip_horizontal
    strength: 1.0
  - operation: flip
    parameter_name: flip_vertical
    strength: 1.0
  - operation: blur
    parameter_name: sigma
    range: [0.5, 2.0]
    steps: 5
    strength: 1.0
    padding_mode: border
  - operation: sharpen
    parameter_name: sharpen_amount
    range: [0.5, 2.0]
    steps: 5
    apply_to_timesteps:
    - "0-2"
    - "7-9"
    - "ALL"
    apply_to_layers:
    - "0-5"
    - "13-18"
    - "24-29"
    - "ALL"
    strength: 1.0
    padding_mode: border
  - operation: amplify
    parameter_name: amplify_factor
    range: [0.5, 1.5]
    steps: 5
    target_token:
    - "ALL"
    - "rose, horse, ship, actor, kiss"
    strength: 1.0
  - operation: translate
    parameter_name: translate_x
    range: [-0.5, 0.05]
    steps: 6
    strength: 1.0
    padding_mode: border
  - operation: translate
    parameter_name: translate_y
    range: [-0.5, 0.05]
    steps: 6
    strength: 1.0
batch_name: dual_gpu_comprehensive_20260129_050427_gpu0
cfg_schedule_settings:
  apply_to_guidance_2: true
  enabled: false
  force_cfg: false
  interpolation: linear
  schedule: {}
  verbose: false
memory_settings:
  clear_cache_between_videos: true
  enable_memory_efficient_attention: true
  enable_memory_optimization: true
  reload_model_for_large_models: false
  use_gradient_checkpointing: true
model_settings:
  cfg_scale: 4.5
  clip_skip: 1
  eta: 0.0
  model_id: Wan-AI/Wan2.1-T2V-1.3B-Diffusers
  sampler: unipc
  seed: 41
  steps: 10
output_dir: outputs/dual_gpu_comprehensive_20260129_050427
prompt_schedule_settings:
  enabled: false
  interpolation: slerp
  schedule: {}
  verbose: false
prompt_settings:
  base_weight: 1.0
  embedding_method: norm_preserving
  enable_prompt_weighting: true
  enable_weighting: false
  negative_prompt: ''
  use_weighted_embeddings: false
  variation_weight: 1.5
template: "[Close-up focused on a single red (rose) in a glass vase | Medium-shot focused on a white (horse) galloping in a grassy (field) | Long-shot A large wood pirate (ship) tossing in stormy ocean (waves) | Long-shot shot of lead (actor), walking directly towards the camera, face forward, dramatic | Medium shot of a romantic hollywood (kiss) between two (people) in love, two faces, cinematic]"
use_timestamp: false
video_settings:
  duration:
  fps: 16
  frames: 25
  height: 368
  width: 640
videos_per_variation: 3
\end{lstlisting}

\end{document}